\documentclass[12pt]{article}
\usepackage{a4wide}
\usepackage{latexsym}
\usepackage{amsmath}
\usepackage{amsfonts}
\usepackage{amscd}
\usepackage{amsthm}
\usepackage{shuffle}
\usepackage{slashed} 
\usepackage{cite}
\usepackage{graphicx}
\usepackage{float}
\usepackage{axodraw}

\usepackage{pslatex}
\usepackage[latin1,utf8]{inputenc}
\usepackage[OT2,T1]{fontenc}

\usepackage{ifthen}

\allowdisplaybreaks

\newcommand{\bq}{\begin{eqnarray}}
\newcommand{\eq}{\end{eqnarray}}
\newcommand{\eps}{\varepsilon}

\begin{document}

\thispagestyle{empty}

\begin{flushright}
  MITP/19-015
\end{flushright}

\vspace{1.5cm}

\begin{center}
  {\Large\bf On the vanishing of certain cuts or residues of loop integrals with higher powers of the propagators\\
  }
  \vspace{1cm}
  {\large Robin Baumeister, Daniel Mediger, Julia Pe\v{c}ovnik and Stefan Weinzierl\\
\vspace{2mm}
      {\small \em PRISMA Cluster of Excellence, Institut f{\"u}r Physik, }\\
      {\small \em Johannes Gutenberg-Universit{\"a}t Mainz,}\\
      {\small \em D - 55099 Mainz, Germany}\\
  } 
\end{center}

\vspace{2cm}

\begin{abstract}\noindent
  {
Starting from two-loops, there are Feynman integrals with higher powers of the propagators.
They arise from self-energy insertions on internal lines.
Within the loop-tree duality approach or within methods based on numerical unitarity one needs (among other things)
the residue when a raised propagator goes on-shell.
We show that for renormalised quantities 
in the on-shell scheme 
these residues can be made to vanish already at the integrand level.
   }
\end{abstract}

\vspace*{\fill}

\newpage

\section{Introduction}
\label{sect:intro}

The aim for theoretical precision predictions for the LHC requires next-to-next-to-leading order (NNLO) calculations for
a number of processes.
If one goes beyond the simplest $2 \rightarrow 2$-processes,
considering a $2 \rightarrow (n-2)$-process with possibly $n>4$, one is in particular interested
in methods which allow for automation.
Numerical methods 
like numerical loop integration \cite{Soper:1998ye,Soper:1999xk,Nagy:2003qn,Gong:2008ww,Assadsolimani:2009cz,Assadsolimani:2010ka,Becker:2010ng,Becker:2011vg,Becker:2012aq,Becker:2012nk,Becker:2012bi,Goetz:2014lla,Seth:2016hmv}
combined with loop-tree duality \cite{Catani:2008xa,Bierenbaum:2010cy,Bierenbaum:2012th,Buchta:2014dfa,Hernandez-Pinto:2015ysa,Buchta:2015wna,Sborlini:2016gbr,Driencourt-Mangin:2017gop,Driencourt-Mangin:2019aix,Runkel:2019yrs}
or methods based on numerical unitarity \cite{Ita:2015tya,Abreu:2017idw,Abreu:2017xsl,Abreu:2017hqn,Abreu:2018jgq}
are a promising path for this approach.

Starting from two-loops, there are Feynman integrals with higher powers of the propagators.
They arise from self-energy insertions on internal lines.
An example is shown in fig.~\ref{fig_1}.
Note that the contributions we are concerned about are not an artefact of a gauge choice.
For gauge theories, we will use Feynman gauge throughout this paper.
In Feynman gauge the Feynman propagator has just simple poles.
In an analytic calculation Feynman integrals with higher powers of the propagators are not a problem.
They are reduced by integration-by-parts identities to master integrals.
The master integrals are then calculated analytically.
It is possible that the set of master integrals itself contains integrals with raised propagators.

The situation is different for numerical approaches.
In this paper we focus on numerical loop integration \cite{Soper:1998ye,Soper:1999xk,Nagy:2003qn,Gong:2008ww,Assadsolimani:2009cz,Assadsolimani:2010ka,Becker:2010ng,Becker:2011vg,Becker:2012aq,Becker:2012nk,Becker:2012bi,Goetz:2014lla,Seth:2016hmv}
in combination with loop-tree duality \cite{Catani:2008xa,Bierenbaum:2010cy,Bierenbaum:2012th,Buchta:2014dfa,Hernandez-Pinto:2015ysa,Buchta:2015wna,Sborlini:2016gbr,Driencourt-Mangin:2017gop,Driencourt-Mangin:2019aix,Runkel:2019yrs}.
Let us mention that our results have also implications for methods
based on numerical unitarity \cite{Ita:2015tya,Abreu:2017idw,Abreu:2017xsl,Abreu:2017hqn,Abreu:2018jgq}.
Raised propagators have been considered previously in
\cite{Bierenbaum:2012th,Sogaard:2014ila,Abreu:2017idw}.
Within these numerical approaches one is interested in
the residue when a raised propagator goes on-shell.
If $f(z)$ is a function of a complex variable $z$, which has a pole of order $\nu$ at $z_0$, the standard formula for the 
residue at $z_0$ is given
by
\bq
\label{residue_example_one_dim}
 \mathrm{res}\left(f,z_0\right)
 & = &
 \frac{1}{\left(\nu-1\right)!}
 \left.
 \left(\frac{d}{dz}\right)^{\nu-1}
 \left[ \left(z-z_0\right)^\nu f\left(z\right) \right]
 \right|_{z=z_0}.
\eq
We may think of the variable $z$ as being the energy flowing through the raised propagator.
For $\nu > 1$ we have a derivative acting on all $z$-dependent quantities in the diagram.
Although this can be done, it is process-dependent and not very well suited for automation.
(Eq.~(\ref{residue_example_one_dim}) is a simple univariate example, the generalisation to the multivariate case
is discussed in ref.~\cite{Sogaard:2014ila}. The computation of the residue in the multivariate case with higher powers of the propagators
is based on Gr\"obner bases.)
Alternatively, ref.~\cite{Bierenbaum:2012th} proposes to reduce Feynman integrals with raised propagators through integration-by-parts
identities to Feynman integrals without raised propagators.
This is possible, but again it is process-dependent and therefore not very well suited for automation.

Although our focus is on the loop-tree duality method, where we cut an $l$-loop contribution exactly $l$-times,
let us briefly comment on the numerical unitarity method.
Here, one writes the $l$-loop amplitude as a linear combination of (known) master integrals with (unknown) coefficients.
The coefficients are determined by cutting the internal propagators, starting with the maximal cut and working down
the hierarchy.
The method exploits the fact that the integrand of a loop amplitude factorises on leading poles into products of tree amplitudes.
However, if higher powers of the propagators are present, one needs in addition to the coefficient of the leading poles
also the coefficients of the subleading poles, where the above mentioned factorisation property no longer holds.
Ref.~\cite{Abreu:2017idw} presents a numerical method to extract the coefficients of the subleading poles by considering equations obtained from cutting less propagators.

Let us now return to the loop-tree duality method.
We would like to isolate the complication into a small process-independent part.
If we only look at the left diagram of fig.~\ref{fig_1} there is nothing we can do.
However, we may look at the set of all diagrams corresponding to a self-energy insertion on a specific internal line.
At two-loops and in $\phi^3$-theory there are two diagrams, as shown in fig.~\ref{fig_1}:
The left diagram of fig.~\ref{fig_1}, which we already discussed, and the right diagram of fig.~\ref{fig_1}, corresponding
to the counterterm from renormalisation.
In the on-shell scheme the counterterm is basically the Taylor expansion to second order around the on-shell value of the self-energy.
Thus, if we would perform the one-loop calculation of the self-energy analytically and combine it with the counterterm, 
we would obtain a transcendental function, which vanishes quadratically in the on-shell limit.
This will cancel the double pole and the residue will vanish.
This is fine, but has the drawback that we introduced transcendental functions from an analytic one-loop calculation.
We would like to work entirely with rational functions, as we do in the numerical approach.
It is therefore natural to ask, if there exists an integral representation for the counterterm, such that the residue
vanishes already at the integrand level.
This is the topic of this paper and we show that such an integral representation for the counterterm exists.
Such integral representations are not unique.
There is quite some freedom to construct an integral representation, only the integral, the UV-behaviour and the on-shell behaviour 
is fixed by the requirement that the counterterm should be a proper counterterm, local in loop momentum space 
and leading to a vanishing residue.
A sufficient condition for the last condition to hold is that the sum of the integrands for the self-energy vanishes quadratically as
the external momentum of the self-energy goes on-shell.
Thus
\bq
\label{eq_condition}
 \lim\limits_{k^2\rightarrow m^2} \left(\mbox{Self-energy integrand}\right)
 & = &
 {\mathcal O}\left( \left(E-E^\flat\right)^2 \right),
\eq
where $E^\flat$ denotes the on-shell value of the energy flowing through the raised propagator, 
This condition will cancel the double pole (and the single pole)
from the propagators, resulting in a vanishing residue.
For gauge theories it is sufficient to require eq.~(\ref{eq_condition}) only up to gauge terms, which vanish when contracted
into gauge-invariant quantities.

In this paper we construct counterterms with the property given in eq.~(\ref{eq_condition}).
Thus, the main result of this paper is that when summed over all relevant diagrams (including counterterms from renormalisation)
residues due to higher poles from self-energy insertions on internal lines can be made to vanish at the integrand level.

Let us mention that the counterterms we construct have higher powers of the propagators 
in the self-energy parts.
At first sight, this may seem like nothing has been gained: 
We removed higher powers of the propagators in one part,
but introduced new higher powers of the propagators in another part.
The essential point is that we removed the higher powers of the propagators from the process-dependent part
and isolated the higher powers of the propagators in a universal process-independent part.
The derivatives for the residues may therefore be calculated once and for all.

One final remark: Although the explicit results for the counterterms
for $\phi^3$-theory and QCD presented in this paper are for stable particles, where the masses are real,
this assumption is not essential.
We may allow complex masses.
We only require that the renormalised propagator has a pole at the renormalised mass with residue $1$.
Our method has a straightforward extension towards the complex mass scheme \cite{Denner:2005fg}.

This paper is organised as follows:
In the next section we consider a simple toy example from complex analysis.
In section~\ref{sect:scalar_theory} we present our argument in detail for the case of a scalar $\phi^3$-theory.
All essential features are already in there.
In section~\ref{sect:QCD} we specialise to the case of quantum chromodynamics, treating spin $1/2$-fermions and massless spin $1$-gauge bosons.
Finally, our conclusions are contained in section~\ref{sect:conclusions}.
Appendix~\ref{sect:Feynman_rules} lists the Feynman rules for the scalar $\phi^3$-theory.

\section{A toy example}
\label{sect:toy}

\begin{figure}
\begin{center}
\includegraphics[scale=1.0]{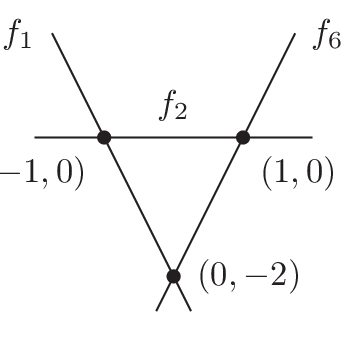}
\hspace*{20mm}
\includegraphics[scale=1.0]{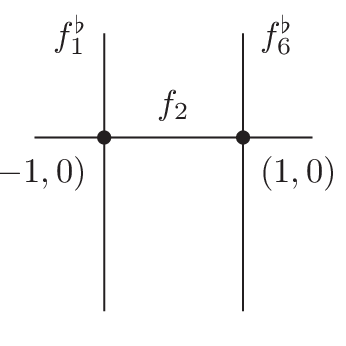}
\end{center}
\caption{
The left figure shows the location of the residues in the $(z_1,z_2)$-plane of the rational function $R$,
the right figure shows the location of the residues of the rational function $R_{\mathrm{CT}}$.
}
\label{fig_toy}
\end{figure}

Let us first look at a toy example and consider the polynomials
\bq
 f_2 \; = \; z_2,
 \;\;\;\;\;\;
 f_1 \; = \; z_1 + \frac{1}{2} z_2 + 1,
 \;\;\;\;\;\;
 f_6 \; = \; z_1 - \frac{1}{2} z_2 - 1
\eq
in two complex variables $z_1$ and $z_2$.
We are interested in the local residues (i.e. two-fold residues in $z_1$ and $z_2$) of the rational function
\bq
 R 
 & = &
 \frac{1}{f_2^2 f_1 f_6}.
\eq
The local residues are at
\bq
 \left(z_1,z_2\right)
 & \in &
 \left\{
 \; \left(-1,0\right),
 \; \left(1,0\right),
 \; \left(0,-2\right) \;
 \right\}.
\eq
The location of the residues is shown in the left drawing of fig.~\ref{fig_toy}.
We are in particular interested in the local residues at $P_1=(-1,0)$ and $P_2=(1,0)$, where we have a double pole from $f_2^2$. We have
\bq
 \mathrm{res}\left(R,P_1\right)
 \; = \;
 \frac{1}{4},
 & &
 \mathrm{res}\left(R,P_2\right)
 \; = \;
 - \frac{1}{4}.
\eq
Let us now define
\bq
 f_1^\flat \; = \; z_1 + 1,
 \;\;\;\;\;\;
 f_6^\flat \; = \; z_1 - 1
\eq
and consider rational functions with poles only along $f_2$, $f_1^\flat$ and $f_6^\flat$, i.e. rational functions
of the form
\bq
 \frac{P\left(z_1,z_2\right)}{f_2^{\nu_2} \left(f_1^\flat\right)^{\nu_1} \left(f_6^\flat\right)^{\nu_6}},
\eq
with $\nu_1,\nu_2,\nu_6 \in {\mathbb N}$ and $P(z_1,z_2)$ a polynomial in $z_1$ and $z_2$.
These functions have local residues only at the two points $P_1=(-1,0)$ and $P_2=(1,0)$
(this is shown in the right picture of fig.~\ref{fig_toy}), and we are interested in a function
$R_{\mathrm{CT}}$ which cancels the residues of $R$ at $P_1$ and $P_2$.
Let us first note that the function
\bq
 R_{\mathrm{try}}
 & = &
 \frac{1}{f_2^2 f_1^\flat f_6^\flat}
 \;\; = \;\;
 \frac{1}{z_2^2 \left(z_1+1\right) \left(z_1-1\right)}
\eq
has no residues at $P_1$ or $P_2$:
\bq
 \mathrm{res}\left(R_{\mathrm{try}},P_1\right)
 \; = \;
 0,
 & &
 \mathrm{res}\left(R_{\mathrm{try}},P_2\right)
 \; = \;
 0,
\eq
since $R_{\mathrm{try}}$ does not have a single pole in $z_2$.
However, expanding $R$ to second order in $z_2$ does the job:
\bq
 R_{\mathrm{CT}}
 & = &
 -
 \frac{1}{f_2^2 f_1^\flat f_6^\flat}
 \left( 1 - \frac{z_2}{2 f_1^\flat} + \frac{z_2}{2 f_6^\flat} \right)
\eq
We have
\bq
 \mathrm{res}\left(R_{\mathrm{CT}},P_1\right)
 \; = \;
 - \frac{1}{4},
 & &
 \mathrm{res}\left(R_{\mathrm{CT}},P_2\right)
 \; = \;
 \frac{1}{4}.
\eq
Thus
\bq
 \mathrm{res}\left(R+R_{\mathrm{CT}},P_1\right)
 \; = \; 
 \mathrm{res}\left(R+R_{\mathrm{CT}},P_2\right)
 & = & 0,
\eq
and the residues at $P_1$ or $P_2$ cancel in the sum.

The analogy with quantum field theory is as follows: We may think of $z_1$ and $z_2$ as two energy
variables, $f_1$, $f_2$ and $f_6$ as propagators and of $f_1^\flat$ and $f_6^\flat$ as the on-shell projections
of $f_1$ and $f_6$, respectively, as $f_2$ goes on-shell.

\section{The method for a scalar theory}
\label{sect:scalar_theory}

\begin{figure}
\begin{center}
\includegraphics[scale=1.0]{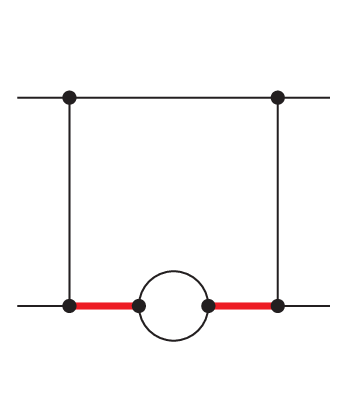}
\hspace*{20mm}
\includegraphics[scale=1.0]{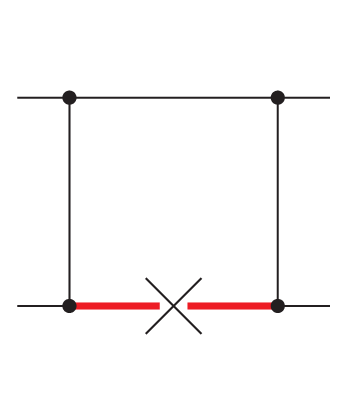}
\end{center}
\caption{
The left figure shows a self-energy insertion on an internal line.
The same momentum is flowing through the two red lines, resulting in a propagator raised to power two.
A self-energy insertion on an internal line is always accompanied by a counterterm, shown in the right figure.
}
\label{fig_1}
\end{figure}

\begin{figure}
\begin{center}
\includegraphics[scale=1.0]{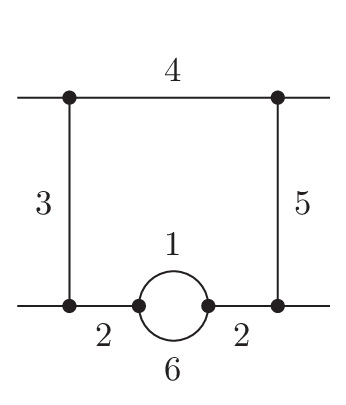}
\hspace*{20mm}
\includegraphics[scale=1.0]{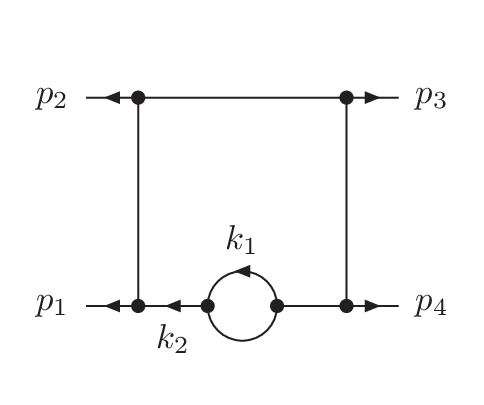}
\end{center}
\caption{
The labelling of the propagators (left figure) and the labelling of the momenta (right figure).
}
\label{fig_2}
\end{figure}

\begin{figure}
\begin{center}
\includegraphics[scale=1.0]{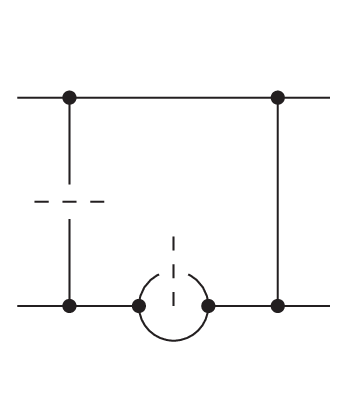}
\hspace*{20mm}
\includegraphics[scale=1.0]{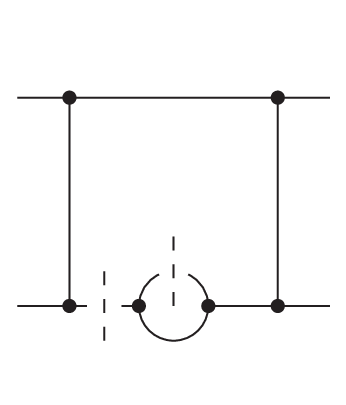}
\hspace*{20mm}
\includegraphics[scale=1.0]{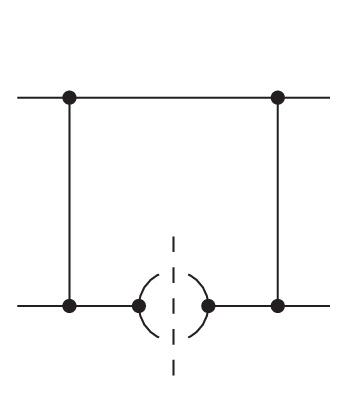}
\end{center}
\caption{
Various cuts of the two-loop diagram, which correspond to spanning trees.
The cut $(1,3)$ (left diagram) is unproblematic.
The cut $(1,2)$ (middle diagram) requires the residue of a doubled propagator.
The right diagram shows the cut $(1,6)$.
}
\label{fig_3}
\end{figure}

Let us now discuss a simple quantum field theory.
We consider a massive $\phi^3$-theory.
The Lagrangian in renormalised quantities is given by
\bq
 {\mathcal L} 
 & = &
 \frac{1}{2} \left( \partial_\mu \phi \right) \left( \partial^\mu \phi \right)
 - \frac{1}{2} m^2 \phi^2
 + \frac{1}{3!} \lambda^{(D)} \phi^3
 +  {\mathcal L}_{\mathrm{CT}}.
\eq
Under renormalisation we have
\bq
 \phi_0 = Z_\phi^{\frac{1}{2}} \phi,
 \;\;\;\;\;\;
 \lambda_0 = Z_\lambda \; \lambda^{(D)},
 \;\;\;\;\;\;
 m_0 = Z_m m,
\eq
where we denote bare quantities with a subscript ``0''.
We work in dimensional regularisation and set $D=4-2\eps$.
We further set
\bq
\label{def_lambda_D}
 \lambda^{(D)}
 & = &
 \mu^\eps S_\eps^{-\frac{1}{2}} \lambda.
\eq
The arbitrary scale $\mu$ is introduced to keep the mass dimension of the renormalised coupling $\lambda$ equal to one.
The factor $S_\eps = (4\pi)^\eps \exp(-\eps \gamma_E)$ absorbs artefacts of dimensional regularisation 
(logarithms of $4\pi$ and Euler's constant $\gamma_E$).
The Lagrangian for the counterterms is given by
\bq
 {\mathcal L}_{\mathrm{CT}}
 & = &
 - \frac{1}{2} \left(Z_\phi-1\right) \phi \Box \phi 
 - \frac{1}{2} \left(Z_\phi Z_m^2 -1 \right) m^2 \phi^2
 + \frac{1}{3!} \left(Z_\phi^{\frac{3}{2}} Z_\lambda - 1 \right) \lambda^{(D)} \phi^3.
\eq
The Feynman rules for the scalar $\phi^3$-theory are listed in appendix~\ref{sect:Feynman_rules}.
For the perturbative expansion of the renormalisation constants we write
\bq
 Z_a
 & = &
 1 + \sum_{n=1}^\infty Z_a^{(n)} \left( \frac{\lambda^2}{\left(4\pi\right)^2} \right)^n,
 \;\;\;\;\;\;
 a \in \{ \phi, m, \lambda \}.
\eq
We will need $Z_m^{(1)}$ and $Z_\phi^{(1)}$.
In the on-shell scheme these renormalisation constants are given by
\bq
\label{renormalisation_constants_phi_cubed}
 Z_m^{(1)}
 & = & 
 \frac{1}{4m^2} B_0\left(m^2,m^2,m^2\right),
 \nonumber \\
 Z_\phi^{(1)}
 & = & 
 \frac{2-\eps}{6m^2} B_0\left(m^2,m^2,m^2\right) - \frac{1-\eps}{3m^4} A_0\left(m^2\right).
\eq
The scalar one-loop integrals $A_0$ and $B_0$ are defined by
\bq
 A_0\left(m^2\right)
 & = &
 16 \pi^2
  S_\eps^{-1} \mu^{2\eps} 
 \int \frac{d^Dk}{(2 \pi)^D i } \; 
 \frac{1}{k^2-m^2},
 \nonumber \\
 B_0\left(p^2,m_1^2,m_2^2\right)
 & = &
 16 \pi^2
  S_\eps^{-1} \mu^{2\eps} 
 \int \frac{d^Dk}{(2 \pi)^D i } \; 
 \frac{1}{\left[\left(k+\frac{1}{2}p\right)^2-m_1^2\right]\left[\left(k-\frac{1}{2}p\right)^2-m_2^2\right]}.
\eq
In this paper we are concerned with diagrams like the one shown in the left picture of fig.~\ref{fig_1}.
In fig.~\ref{fig_2} we show our choice for the labelling of the propagators and the orientation of the momenta.
With
\bq
 D_j & = & k_j^2 - m^2 + i \delta
\eq
we have for this diagram
\bq
 I_{\mathrm{twoloop}}
 & = &
 \frac{i \lambda^6}{2}
 \mu^{4\eps} S_\eps^{-2}
 \int \frac{d^Dk_1}{\left(2\pi\right)^D}
 \int \frac{d^Dk_2}{\left(2\pi\right)^D}
 \frac{1}{D_1 D_2^2 D_3 D_4 D_5 D_6},
\eq
where we ignored a prefactor $\mu^{2\eps} S_\eps^{-1}$, accompanying also the Born amplitude.
We see that $D_2$ is raised to the power two.
Within the loop-tree duality method we take residues in the energy integrations $E_1$ and $E_2$.
The residues are classified by the set of spanning trees for our diagram.
We may denote a spanning tree by the propagators we remove to get a tree diagram.
The set of spanning trees for our two-loop diagram is given by
\bq
 \left\{
 \;
  \left(1,2\right),
 \;
  \left(1,3\right),
 \;
  \left(1,4\right),
 \;
  \left(1,5\right),
 \;
  \left(1,6\right),
 \;
  \left(2,6\right),
 \;
  \left(3,6\right),
 \;
  \left(4,6\right),
 \;
  \left(5,6\right)
 \;
 \right\}.
\eq
Each spanning tree defines also a cut graph.
For a cut graph, we don't remove internal edges but cut them into half-edges.
The half-edges become additional external lines of the cut graph.
In fig.~\ref{fig_3} we show a few examples of cut graphs obtained from spanning trees. 
Problematic are the cuts $(1,2)$ and $(2,6)$.
As $D_2$ occurs quadratically, taking the residue for $D_2=0$ forces us to compute a derivative.

In this paper we would like to point out, that the left diagram of fig.~\ref{fig_1} always comes 
in combination with a counterterm, shown in the right picture of fig.~\ref{fig_1}.
The contribution from the counterterm is 
\bq
 I_{\mathrm{twoloop},\mathrm{CT}}
 & = &
 -
 \frac{\lambda^6}{\left(4\pi\right)^2}
 \mu^{2\eps} S_\eps^{-1}
 \int \frac{d^Dk_2}{\left(2\pi\right)^D}
 \frac{\left[Z_\phi^{(1)} k_2^2 - \left(Z_\phi^{(1)} + 2 Z_m^{(1)}\right) m^2 \right]}{D_2^2 D_3 D_4 D_5}.
\eq
Let us write
\bq
 I_{\mathrm{twoloop}}
 & = &
 i \lambda^6
 \mu^{4\eps} S_\eps^{-2}
 \int \frac{d^Dk_1}{\left(2\pi\right)^D}
 \int \frac{d^Dk_2}{\left(2\pi\right)^D}
 R_{\mathrm{twoloop}}\left(k_1,k_2\right),
 \nonumber \\
 R_{\mathrm{twoloop}}\left(k_1,k_2\right)
 & = &
 \frac{1}{2 D_1 D_2^2 D_3 D_4 D_5 D_6}.
\eq
$R_{\mathrm{twoloop}}(k_1,k_2)$ is a rational function in $k_1$ and $k_2$.
Within the numerical method one writes 
$I_{\mathrm{twoloop},\mathrm{CT}}$ also as a two-loop integral:
\bq
\label{twoloop_CT}
 I_{\mathrm{twoloop},\mathrm{CT}}
 & = &
 i \lambda^6
 \mu^{4\eps} S_\eps^{-2}
 \int \frac{d^Dk_1}{\left(2\pi\right)^D}
 \int \frac{d^Dk_2}{\left(2\pi\right)^D}
 R_{\mathrm{twoloop},\mathrm{CT}}\left(k_1,k_2\right).
\eq
We may now ask the question if there exists a function $R_{\mathrm{twoloop},\mathrm{CT}}(k_1,k_2)$, rational
in the energies $E_1$ and $E_2$, such that
\begin{enumerate}
\item $R_{\mathrm{twoloop},\mathrm{CT}}(k_1,k_2)$ satisfies eq.~(\ref{twoloop_CT}),
\item the sum of $R_{\mathrm{twoloop}}$ and $R_{\mathrm{twoloop},\mathrm{CT}}$ falls off 
for $|k_1|\rightarrow \infty$ as $|k_1|^{-5}$, i.e.
\bq
 \lim\limits_{|k_1|\rightarrow \infty} \left( R_{\mathrm{twoloop}}\left(k_1,k_2\right) + R_{\mathrm{twoloop},\mathrm{CT}}\left(k_1,k_2\right) \right)
 & = & {\mathcal O}\left(|k_1|^{-5}\right),
\eq
\item the sum of $R_{\mathrm{twoloop}}$ and $R_{\mathrm{twoloop},\mathrm{CT}}$
vanishes quadratically as $k_2$ goes on-shell, i.e.
\bq
 \lim\limits_{k_2^2 \rightarrow m^2} \left( R_{\mathrm{twoloop}}\left(k_1,k_2\right) + R_{\mathrm{twoloop},\mathrm{CT}}\left(k_1,k_2\right) \right)
 & = & {\mathcal O}\left( \left(E_2-E_2^\flat\right)^2 \right),
\eq
\item $R_{\mathrm{twoloop},\mathrm{CT}}(k_1,k_2)$ is independent of the energy $E_2$.
\end{enumerate}
The first two requirements are just the statement that $R_{\mathrm{twoloop},\mathrm{CT}}$ is 
a local counterterm at the integrand level 
for the ultraviolet sub-divergence given by the self-energy sub-graph.
Requirement 3 is the new condition which we would like to enforce and ensures that the residue
from $D_2 \rightarrow 0$ will vanish.
Condition 4 is an additional technical requirement and ensures that 
$I_{\mathrm{twoloop},\mathrm{CT}}$ does not receive contributions from the cut $(1,6)$.
This cut is shown
in the right diagram of fig.~\ref{fig_3}.

Let us point out that all conditions laid out above
refer only to the self-energy sub-diagram, not to the full diagram.
The conditions are therefore universal process-independent conditions.

\begin{figure}
\begin{center}
\includegraphics[scale=1.0]{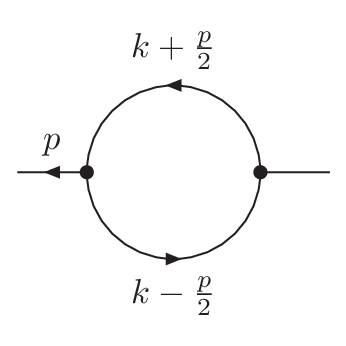}
\end{center}
\caption{
The labelling of the momenta for the one-loop self-energy.
}
\label{fig_4}
\end{figure}

Let us now look at the self-energy.
It is convenient to adopt a slightly different notation for the momenta, shown in fig.~\ref{fig_4}.
For the (bare) one-loop self-energy
we have
\bq
 -i \Sigma_{\mathrm{oneloop}}
 & = &
 \lambda^2 \mu^{2\eps} S_\eps^{-1}
 \int \frac{d^Dk}{\left(2\pi\right)^D}
 R_{\mathrm{oneloop}},
 \;\;\;\;\;\;
 R_{\mathrm{oneloop}}
 \; = \;
 \frac{1}{2 D_1 D_2},
 \nonumber \\
 & &
 D_1
 \; = \; \left(k+\frac{1}{2}p\right)^2 - m^2,
 \;\;\;\;\;\;
 D_2
 \; = \; \left(k-\frac{1}{2}p\right)^2 - m^2.
\eq
Given $p=(E,\vec{p})$ we define $p^\flat$ by
\bq
\label{def_p_flat}
 p^\flat & = & \left( \mathrm{sign}\left(E\right) \sqrt{\vec{p}^2+m^2} ,\vec{p}\right).
\eq
The momentum $p^\flat$ is on-shell
\bq
 \left(p^\flat\right)^2 & = & m^2,
\eq
and does not depend on $E$ (apart from the sign).
With $n=(1,\vec{0})$ we may write $p^\flat$ equally as
\bq
 p^\flat
 & = &
 p - c n,
 \;\;\;\;\;\;
 c \; = \;
 \frac{1}{2n^2} \left( 2 p \cdot n - \mathrm{sign}\left(2 p \cdot n\right) \sqrt{ \left(2 p \cdot n\right)^2 - 4 n^2 \left(p^2-m^2\right)} \right).
\eq
For the counterterm we write
\bq
 -i \Sigma_{\mathrm{oneloop},\mathrm{CT}}
 & = &
 \lambda^2 \mu^{2\eps} S_\eps^{-1}
 \int \frac{d^Dk}{\left(2\pi\right)^D}
 R_{\mathrm{oneloop},\mathrm{CT}}.
\eq
We require that the only poles of $R_{\mathrm{oneloop},\mathrm{CT}}$ originate from
\bq
 D_1^\flat 
 \; = \;
 \left(k + \frac{1}{2}p^\flat\right)^2 - m^2,
 \;\;\;\;\;\;
 D_2^\flat 
 \; = \;
 \left(k - \frac{1}{2}p^\flat\right)^2 - m^2.
\eq
$D_1^\flat$ and $D_2^\flat$ are the images of $D_1$ and $D_2$ under the map $p \rightarrow p^\flat$.
A possible choice for $R_{\mathrm{oneloop},\mathrm{CT}}$ is given by
\bq
\label{counterterm_phi_cubed}
 R_{\mathrm{oneloop},\mathrm{CT}}
 & = &
 - \frac{1}{2 D_1^\flat D_2^\flat}
 \left[ 1 - \frac{4 k \cdot \left(p-p^\flat\right) + p^2-m^2}{4 D_1^\flat} + \frac{4 k \cdot \left(p-p^\flat\right) - p^2+m^2}{4 D_2^\flat} \right]
 \nonumber \\
 & &
 + \frac{\left(p-p^\flat\right)^2}{8 m^2} 
   \left( \frac{2}{D_1^\flat D_2^\flat} - \frac{1}{\left(D_1^\flat\right)^2} - \frac{1}{\left(D_2^\flat\right)^2} \right).
\eq
The first line is the expansion of $R_{\mathrm{oneloop}}$ around the on-shell kinematics, such that
the difference between the first line and $R_{\mathrm{oneloop}}$ is of order ${\mathcal O}((p^2-m^2)^2)$.
The first line gives also a local UV-counterterm, such that
the difference between the first line and $R_{\mathrm{oneloop}}$ is of order ${\mathcal O}(|k|^{-5})$ or better.
Thus, we see that the first line satisfies conditions~(2) and (3).
Condition~(4) is trivially satisfied due to our definition of $p^\flat$ in eq.~(\ref{def_p_flat}).
It remains to satisfy condition~(1).
This is the job of the term in the second line.
This term ensures that
\bq
 \lambda^2 \mu^{2\eps} S_\eps^{-1}
 \int \frac{d^Dk}{\left(2\pi\right)^D}
 R_{\mathrm{oneloop},\mathrm{CT}}
 & = &
 \frac{\lambda^2}{\left(4\pi\right)^2}
 \;
 i \left[ Z_\phi^{(1)} p^2 - \left( Z_\phi^{(1)} + 2 Z_m^{(1)} \right) m^2 \right],
\eq
where $Z_\phi^{(1)}$ and $Z_m^{(1)}$ have been defined in eq.~(\ref{renormalisation_constants_phi_cubed}).
At the same time, the term on the second line does not spoil the on-shell limit nor the UV-limit.
To see this, we note that
\bq
 \left(p-p^\flat\right)^2
\eq
vanishes quadratically in the on-shell limit.
Secondly, the combination
\bq
 \frac{2}{D_1^\flat D_2^\flat} - \frac{1}{\left(D_1^\flat\right)^2} - \frac{1}{\left(D_2^\flat\right)^2}
\eq
falls off as $\mathcal{O}(|k|^{-6})$ in the UV-limit.

The counterterm $R_{\mathrm{oneloop},\mathrm{CT}}$ is a rational function in the energy variable $E_k$.
An inspection of eq.~(\ref{counterterm_phi_cubed}) shows that $R_{\mathrm{oneloop},\mathrm{CT}}$ has double poles in the variable $E_k$.
This is however unproblematic, as it occurs in a universal building block. The residues can be calculated once and for all.
As an example we consider the residue at
\bq
 E_{k,D_1^\flat} & = & - \frac{1}{2} E_{p^\flat} + \sqrt{\left(\vec{k}+\frac{1}{2}\vec{p}\right)^2+m^2}.
\eq
As an abbreviation we set
\bq
 E_1 & = &  \sqrt{\left(\vec{k}+\frac{1}{2}\vec{p}\right)^2+m^2}.
\eq
We find
\bq
 \mathrm{res}\left(R_{\mathrm{oneloop},\mathrm{CT}}, E_k = E_{k,D_1^\flat}\right)
 & = &
- \frac{1}{4 E_1 D_2^\flat}
 + \frac{\left(E_p-E_{p^\flat}\right)^2}{8 E_1 m^2 D_2^\flat}
 + \frac{\left(E_p-E_{p^\flat}\right)^2}{32 E_1^3 m^2}
 - \frac{\left(E_p-E_{p^\flat}\right)^2}{32 E_1^3 D_2^\flat}
 \nonumber \\
 & &
 - \frac{\left(E_1-E_{p^\flat}\right)\left(E_p-E_{p^\flat}\right)}{2 E_1 \left(D_2^\flat\right)^2}
 + \frac{E_{p^\flat} \left(E_p-E_{p^\flat}\right)^2}{16 E_1^2 \left(D_2^\flat\right)^2},
\eq
where $D_2^\flat$ is understood to be evaluated at $k=(E_{k,D_1^\flat},\vec{k})$.

The rational function $R_{\mathrm{oneloop}}$ has a corresponding residue at
\bq
 E_{k,D_1} & = & - \frac{1}{2} E_{p} + \sqrt{\left(\vec{k}+\frac{1}{2}\vec{p}\right)^2+m^2}.
\eq
$R_{\mathrm{oneloop}}$ has only single poles and the residue is given by
\bq
 \mathrm{res}\left(R_{\mathrm{oneloop}}, E_k = E_{k,D_1}\right)
 & = &
 \frac{1}{4 E_1 D_2},
\eq
where $D_2$ is understood to be evaluated at $k=(E_{k,D_1},\vec{k})$.
For the sum of the two residues we have
\bq
\lefteqn{
 \mathrm{res}\left(R_{\mathrm{oneloop}}, E_k = E_{k,D_1}\right)
 +
 \mathrm{res}\left(R_{\mathrm{oneloop},\mathrm{CT}}, E_k = E_{k,D_1^\flat}\right)
 = } & &
 \nonumber \\
 & &
 \frac{1}{4 E_1}
 \left[
 \frac{1}{D_2}
 - \frac{1}{D_2^\flat}
 - \frac{2 \left(E_1-E_{p^\flat}\right)\left(E_p-E_{p^\flat}\right)}{\left(D_2^\flat\right)^2}
 \right]
 \nonumber \\
 & &
 + \frac{\left(E_p-E_{p^\flat}\right)^2}{8 E_1 m^2 D_2^\flat}
 + \frac{\left(E_p-E_{p^\flat}\right)^2}{32 E_1^3 m^2}
 - \frac{\left(E_p-E_{p^\flat}\right)^2}{32 E_1^3 D_2^\flat}
 + \frac{E_{p^\flat} \left(E_p-E_{p^\flat}\right)^2}{16 E_1^2 \left(D_2^\flat\right)^2}.
\eq
We note that the term in the square bracket vanishes also quadratically in the on-shell limit.
In technical terms we have for $D_2$ evaluated at $k=(E_{k,D_1},\vec{k})$ and for $D_2^\flat$ evaluated at $k=(E_{k,D_1^\flat},\vec{k})$:
\bq
 \frac{1}{D_2}
 - \frac{1}{D_2^\flat}
 - \frac{2 \left(E_1-E_{p^\flat}\right)\left(E_p-E_{p^\flat}\right)}{\left(D_2^\flat\right)^2}
 & = &
 {\mathcal O}\left( \left(E_p-E_{p^\flat}\right)^2 \right).
\eq
Let us now go back to fig.~\ref{fig_1}.
We combine the two-loop diagram (left diagram in fig.~\ref{fig_1})
with the one-loop diagram with a counterterm insertion (right diagram in fig.~\ref{fig_1}).
For the latter we derived a two-loop integral representation.
We may evaluate the sum of the two-loop integrals by taking residues in the two energy integrations.
Our construction ensures that there is no residue from the cut $(1,2)$ (middle diagram of fig.~\ref{fig_3}).
There are of course residues from an unproblematic cut like $(1,3)$ (left diagram of fig.~\ref{fig_3}).
Finally, let us note that the residue for the cut $(1,6)$ (right diagram of fig.~\ref{fig_3})
receives only a contribution from the genuine two-loop diagram, but not from the diagram with the counterterm insertion.
By construction, the integral representation of the counterterm is independent of the energy flowing through
the outer loop, therefore there is no residue in this energy variable.

\section{QCD}
\label{sect:QCD}

Let us now consider QCD with $N_f$ massless quarks and $N_Q$ massive quarks.
It is sufficient to discuss the case where all massive quarks have the same mass $m$.
We denote the renormalisation constant for the gluon field by $Z_3$, 
the one for a massless quark field by $Z_2$ and the one for a massive quark field by $Z_{2,Q}$.
The renormalisation constant for the heavy quark mass $m$ is denoted by $Z_m$. 
For the renormalisation constants we write
\bq
 Z_a
 & = &
 1 + \sum_{n=1}^\infty Z_a^{(n)} \left( \frac{\alpha_s}{4\pi} \right)^n.
\eq
We will need the one-loop renormalisation constants. For $Z_3^{(1)}$ we write
\bq
 Z_3^{(1)}
 & = &
 Z_{3,l}^{(1)}
 + 
 Z_{3,Q}^{(1)},
\eq
separating the contributions from the massless particles in the loop ($Z_{3,l}^{(1)}$) 
from the contribution of the massive quark in the loop ($Z_{3,Q}^{(1)}$).
In the on-shell scheme we have
\bq
 Z_2^{(1)}
 & = &
 0,
 \nonumber \\
 Z_{2,Q}^{(1)}
 & = &
 - \left(3-2\eps\right) C_F B_0\left(m^2,m^2,0\right),
 \nonumber \\
 Z_m^{(1)}
 & = &
 - \left(3-2\eps\right) C_F B_0\left(m^2,m^2,0\right),
 \nonumber \\
 Z_{3,l}^{(1)}
 & = &
 0,
 \nonumber \\
 Z_{3,Q}^{(1)}
 & = &
 - \frac{4}{3} T_R N_Q B_0\left(0,m^2,m^2\right).
\eq
The self-energies are diagonal in colour space. 
We suppress the Kronecker delta's in colour space.

\subsection{Light quarks}

In this paragraph we set
\bq
 D_1^\flat 
 \; = \;
 \left(k + \frac{1}{2}p^\flat\right)^2,
 \;\;\;\;\;\;
 D_2^\flat 
 \; = \;
 \left(k - \frac{1}{2}p^\flat\right)^2.
\eq
The self-energy for a massless quark is given by
\bq
 - i \Sigma_{\mathrm{oneloop}}
 & = & 
 g^2 \mu^{2\eps} S_\eps^{-1} 
 \int \frac{d^Dk}{(2\pi)^D} 
 R_{\mathrm{oneloop}},
 \;\;\;\;\;\;
 R_{\mathrm{oneloop}}
 \; = \;
 C_F
 \frac{2 \left(1-\eps\right) \left( \slashed{k} + \frac{1}{2} \slashed{p} \right)}{D_1 D_2}.
\eq
For the counterterm we write
\bq
 - i \Sigma_{\mathrm{oneloop},\mathrm{CT}}
 & = & 
 g^2 \mu^{2\eps} S_\eps^{-1} 
 \int \frac{d^Dk}{(2\pi)^D} 
 R_{\mathrm{oneloop},\mathrm{CT}}.
\eq
A possible choice for $R_{\mathrm{oneloop},\mathrm{CT}}$ is given by
\bq
 R_{\mathrm{oneloop},\mathrm{CT}}
 & = &
 - C_F \frac{2 \left(1-\eps\right) \left( \slashed{k} + \frac{1}{2} \slashed{p} \right)}{D_1^\flat D_2^\flat}
 \left[ 1 - \frac{4 k \cdot \left(p-p^\flat\right) + p^2}{4 D_1^\flat} + \frac{4 k \cdot \left(p-p^\flat\right) - p^2}{4 D_2^\flat} \right].
 \;\;\;
\eq
Integration is in this case particularly simple. All integrals are scaleless integrals, which vanish in dimensional regularisation. Therefore
\bq
 g^2 \mu^{2\eps} S_\eps^{-1}
 \int \frac{d^Dk}{\left(2\pi\right)^D}
 R_{\mathrm{oneloop},\mathrm{CT}}
 & = &
 \frac{\alpha_s}{4\pi} 
 \;
 i Z_2^{(1)} \slashed{p}
 \; = \; 0.
\eq

\subsection{Heavy quarks}

In this paragraph we set
\bq
 D_1^\flat 
 \; = \;
 \left(k + \frac{1}{2}p^\flat\right)^2 - m^2,
 \;\;\;\;\;\;
 D_2^\flat 
 \; = \;
 \left(k - \frac{1}{2}p^\flat\right)^2.
\eq
The self-energy for a massive quark is given by
\bq
 - i \Sigma_{\mathrm{oneloop}}
 & = & 
 g^2 \mu^{2\eps} S_\eps^{-1} 
 \int \frac{d^Dk}{(2\pi)^D} 
 R_{\mathrm{oneloop}},
 \nonumber \\
 R_{\mathrm{oneloop}}
 & = &
 C_F
 \frac{2 \left(1-\eps\right) \left( \slashed{k} + \frac{1}{2} \slashed{p} \right) - 4 \left(1-\frac{1}{2} \eps \right) m}{D_1 D_2}.
\eq
For the counterterm we write
\bq
 - i \Sigma_{\mathrm{oneloop},\mathrm{CT}}
 & = & 
 g^2 \mu^{2\eps} S_\eps^{-1} 
 \int \frac{d^Dk}{(2\pi)^D} 
 R_{\mathrm{oneloop},\mathrm{CT}}.
\eq
A possible choice for $R_{\mathrm{oneloop},\mathrm{CT}}$ is given by
\bq
\label{counterterm_heavy_quark}
\lefteqn{
 R_{\mathrm{oneloop},\mathrm{CT}}
 = 
 C_F \left\{
 - \frac{2 \left(1-\eps\right) \left( \slashed{k} + \frac{1}{2} \slashed{p}^\flat \right) - 4 \left(1-\frac{1}{2} \eps \right) m}{D_1^\flat D_2^\flat}
 \left[ 1 - \frac{4 k \cdot \left(p-p^\flat\right) + p^2 - m^2}{4 D_1^\flat} 
 \right. \right.
 } & & \nonumber \\
 & &
 \left.
 + \frac{4 k \cdot \left(p-p^\flat\right) - p^2 + m^2}{4 D_2^\flat} \right]
 - \frac{\left(1-\eps\right) \left( \slashed{p} - \slashed{p}^\flat\right)}{D_1^\flat D_2^\flat}
 \nonumber \\
 & &
 - \frac{1}{4} \left( \slashed{p}^\flat - m \right) \left(p^2-m^2\right) \frac{D_1^\flat-D_2^\flat+4m^2}{\left(D_1^\flat\right)^2 \left(D_2^\flat\right)^2}
 + \frac{m \left(p-p^\flat\right)^2}{4 m^2} 
   \frac{\left(D_1^\flat-D_2^\flat\right)\left(D_1^\flat-D_2^\flat+2m^2\right)}{\left(D_1^\flat\right)^2 \left(D_2^\flat\right)^2}
 \nonumber \\
 & &
 + \frac{\left( \slashed{p}^\flat - m \right) \left[ p^\flat \cdot \left(p-p^\flat\right) \right]}{m^2}
   \frac{\left(D_1^\flat-D_2^\flat\right)\left(D_1^\flat-D_2^\flat+\frac{3}{2}m^2\right)}
        {\left(D_1^\flat\right)^2 \left(D_2^\flat\right)^2}
 - \frac{\eps m \left(p-p^\flat\right)^2}{2 \left(D_1^\flat\right)^2 D_2^\flat}
 \nonumber \\
 & &
 \left.
 + \frac{\left[ 2 \left( \slashed{p} - m \right) m^2 - m \left(p^2-m^2\right) \right]}{2m^2}  
   \frac{\left(D_1^\flat-D_2^\flat\right)\left(2 D_1^\flat+D_2^\flat\right)}{\left(D_1^\flat\right)^2 \left(D_2^\flat\right)^2}
 \right\}.
\eq
The terms in the first two lines approximate $R_{\mathrm{oneloop}}$ in the on-shell and in the ultraviolet
limit.
The terms in the third to fifth line ensure that the integration of $R_{\mathrm{oneloop},\mathrm{CT}}$
gives the desired result.
We have
\bq
 g^2 \mu^{2\eps} S_\eps^{-1}
 \int \frac{d^Dk}{\left(2\pi\right)^D}
 R_{\mathrm{oneloop},\mathrm{CT}}
 & = &
 \frac{\alpha_s}{4\pi} 
 \;
 i \left[ Z_2^{(1)} \slashed{p} - \left( Z_2^{(1)} + Z_m^{(1)} \right) m \right].
\eq
The terms in the third to fifth line vanish
in the on-shell and in the ultraviolet
limit.
For example, the last term in eq.~(\ref{counterterm_heavy_quark})
falls of like ${\mathcal O}(|k|^{-5})$ in the UV-limit.
For the on-shell limit we note that
\bq
 2 \left( \slashed{p} - m \right) m^2 - m \left(p^2-m^2\right)
 & = &
 - m \left( \slashed{p} - m \right) \left( \slashed{p} - m \right).
\eq

\subsubsection{The $\overline{\mathrm{MS}}$-scheme}

For the mass of a heavy quark, the $\overline{\mathrm{MS}}$-scheme and the on-shell scheme are two popular
renormalisation schemes.
In this paragraph, we comment on the $\overline{\mathrm{MS}}$-scheme.
In the previous section we constructed an integral representation
$R_{\mathrm{oneloop},\mathrm{CT}}$ in the on-shell scheme with the property
that
\bq
 \lim\limits_{k^2\rightarrow m^2} \left(R_{\mathrm{oneloop}}+R_{\mathrm{oneloop},\mathrm{CT}}\right)
 & = &
 {\mathcal O}\left( \left(E-E^\flat\right)^2 \right).
\eq
This is not possible in the $\overline{\mathrm{MS}}$-scheme.
To see this, let us perform a finite renormalisation from the on-shell mass to the $\overline{\mathrm{MS}}$-mass.
This amounts to adding the term
\bq
\lefteqn{
 \frac{\alpha_s}{4\pi} 
 \;
 i \left\{ 
    \left[ Z_2^{(1)} \slashed{p} - \left( Z_2^{(1)} + Z_{m,\overline{\mathrm{MS}}}^{(1)} \right) m \right]
    -
    \left[ Z_2^{(1)} \slashed{p} - \left( Z_2^{(1)} + Z_m^{(1)} \right) m \right]
   \right\}
 = }
 \nonumber \\
 & = & 
 \frac{\alpha_s}{4\pi} 
 \;
 i \left(
         Z_m^{(1)} - Z_{m,\overline{\mathrm{MS}}}^{(1)}
   \right) m
 \; = \;
 \frac{\alpha_s}{4\pi} 
 \;
 i C_F \left( -4 + 3 \ln\frac{m^2}{\mu^2} \right) m
 + {\mathcal O}\left(\eps\right),
\eq
where we used
\bq
 Z_{m,\overline{\mathrm{MS}}}^{(1)}
 & = &
 - \frac{3 C_F}{\eps}.
\eq
The term from the finite renormalisation is a non-zero constant in the on-shell limit and hence does not vanish
quadratically in the on-shell limit.
Neither can there be an integral representation, which vanishes quadratically in the on-shell limit.

\subsection{Gluons}

We now consider the gluon self-energy.
Let us first briefly discuss what happens in an analytic calculation.
We denote by $-i\Pi^{\mu\nu}_{\mathrm{oneloop}}$ the one-loop contribution to the gluon self-energy 
and by $-i\Pi^{\mu\nu}_{\mathrm{oneloop},\mathrm{CT}}$ the contribution from the counterterm.
The self-energy is transverse and we may write
\bq
 -i \Pi^{\mu\nu}_{\mathrm{oneloop}}
 & = &
  i \left( p^2 g^{\mu\nu} - p^\mu p^\nu \right) \Pi_{\mathrm{oneloop}}\left(p^2\right),
\eq
with a scalar function $\Pi_{\mathrm{oneloop}}(p^2)$.
We expand $\Pi_{\mathrm{oneloop}}(p^2)$ around $p^2=0$:
\bq
 \Pi_{\mathrm{oneloop}}\left(p^2\right)
 & = &
 \Pi_{\mathrm{oneloop}}\left(0\right)
 + {\mathcal O}\left(p^2\right).
\eq
This defines $Z_3^{(1)}$:
\bq
 \frac{\alpha_s}{4\pi} \; Z_3^{(1)} & = & \Pi_{\mathrm{oneloop}}\left(0\right).
\eq
Thus
\bq
 -i \left( \Pi^{\mu\nu}_{\mathrm{oneloop}} + \Pi^{\mu\nu}_{\mathrm{oneloop},\mathrm{CT}} \right)
 & = &
  i \left( p^2 g^{\mu\nu} - p^\mu p^\nu \right) \cdot {\mathcal O}\left(p^2\right).
\eq
We see that the term proportional to $g^{\mu\nu}$ has a factor $(p^2)^2$ and will cancel a double pole from the propagators.
On the other hand, the term proportional to $p^\mu p^\nu$ comes only with a single factor $p^2$, leaving a residue from a single pole.
However, this term is proportional to $p^\mu p^\nu$. We may neglect the contribution from this residue if we contract
this term into quantities, which vanish when contracted with an on-shell momentum $p^\mu$ or $p^\nu$.

For the gluon self-energy we distinguish the case of massless particles in the loop and the case of a massive quark loop.

\subsubsection{Contributions from massless particles}

In this paragraph we set
\bq
 D_1^\flat 
 \; = \;
 \left(k + \frac{1}{2}p^\flat\right)^2,
 \;\;\;\;\;\;
 D_2^\flat 
 \; = \;
 \left(k - \frac{1}{2}p^\flat\right)^2.
\eq
The contribution to the gluon self-energy from massless particles is given by
\bq
 - i \Pi^{\mu\nu}_{\mathrm{oneloop}}
 & = & 
 g^2 \mu^{2\eps} S_\eps^{-1} 
 \int \frac{d^Dk}{(2\pi)^D} 
 R_{\mathrm{oneloop}},
 \nonumber \\
 R_{\mathrm{oneloop}}
 & = &
 \left\{ - 2 C_A \left[ -p^2 g^{\mu\nu} + p^\mu p^\nu - 2 \left( 1 - \eps \right) k^\mu k^\nu 
                      + \frac{1}{2} \left( 1 - \eps \right) g^{\mu\nu} \left( D_1 + D_2 \right) \right]
 \right. \nonumber \\
 & & \left.
       - 2 T_R N_f \left[ p^2 g^{\mu\nu} - p^\mu p^\nu + 4 k^\mu k^\nu 
                          - g^{\mu\nu} \left( D_1 + D_2 \right) \right] \right\}
 \frac{1}{D_1 D_2}.
\eq
For the counterterm we write
\bq
 - i \Pi^{\mu\nu}_{\mathrm{oneloop},\mathrm{CT}}
 & = & 
 g^2 \mu^{2\eps} S_\eps^{-1} 
 \int \frac{d^Dk}{(2\pi)^D} 
 R_{\mathrm{oneloop},\mathrm{CT}}.
\eq
A possible choice for $R_{\mathrm{oneloop},\mathrm{CT}}$ is given by
\bq
\lefteqn{
 R_{\mathrm{oneloop},\mathrm{CT}}
 = 
 \left\{ 2 C_A \left[ -p^2 g^{\mu\nu} + p^\mu p^\nu - 2 \left( 1 - \eps \right) k^\mu k^\nu 
                      + \frac{1}{2} \left( 1 - \eps \right) g^{\mu\nu} \left( D_1 + D_2 \right) \right]
 \right. } & &
 \nonumber \\
 & & \left.
       + 2 T_R N_f \left[ p^2 g^{\mu\nu} - p^\mu p^\nu + 4 k^\mu k^\nu 
                          - g^{\mu\nu} \left( D_1 + D_2 \right) \right] \right\}
 \frac{1}{D_1^\flat D_2^\flat}
 \left\{ 1 - \frac{4 k \cdot \left(p-p^\flat\right) + p^2}{4 D_1^\flat} 
 \right. \nonumber \\
 & & \left. 
 + \frac{4 k \cdot \left(p-p^\flat\right) - p^2}{4 D_2^\flat} 
 + \left[ k \cdot \left(p-p^\flat\right) \right]^2 \left( \frac{1}{\left(D_1^\flat\right)^2} + \frac{1}{\left(D_2^\flat\right)^2} - \frac{1}{D_1^\flat D_2^\flat}\right)
 \right\}.
\eq
Integration yields
\bq
 g^2 \mu^{2\eps} S_\eps^{-1}
 \int \frac{d^Dk}{\left(2\pi\right)^D}
 R_{\mathrm{oneloop},\mathrm{CT}}
 & = &
 \frac{\alpha_s}{4\pi} 
 \;
 i Z_{3,l}^{(1)} \left( - g^{\mu\nu} p^2 + p^\mu p^\nu \right)
 \; = \; 
 0.
\eq

\subsubsection{Contributions from a massive quark}

In this paragraph we set
\bq
 D_1^\flat 
 \; = \;
 \left(k + \frac{1}{2}p^\flat\right)^2 - m^2,
 \;\;\;\;\;\;
 D_2^\flat 
 \; = \;
 \left(k - \frac{1}{2}p^\flat\right)^2 - m^2.
\eq
The contribution to the gluon self-energy from massive quarks is given by
\bq
 - i \Pi^{\mu\nu}_{\mathrm{oneloop}}
 & = & 
 g^2 \mu^{2\eps} S_\eps^{-1} 
 \int \frac{d^Dk}{(2\pi)^D} 
 R_{\mathrm{oneloop}},
 \nonumber \\
 R_{\mathrm{oneloop}}
 & = &
 - 2 T_R N_Q
 \left[
        p^2 g^{\mu\nu} - p^\mu p^\nu + 4 k^\mu k^\nu - g^{\mu\nu} \left( D_1 + D_2 \right) 
 \right]
 \frac{1}{D_1 D_2}.
\eq
For the counterterm we write
\bq
 - i \Pi^{\mu\nu}_{\mathrm{oneloop},\mathrm{CT}}
 & = & 
 g^2 \mu^{2\eps} S_\eps^{-1} 
 \int \frac{d^Dk}{(2\pi)^D} 
 R_{\mathrm{oneloop},\mathrm{CT}}.
\eq
A possible choice for $R_{\mathrm{oneloop},\mathrm{CT}}$ is given by
\bq
\lefteqn{
 R_{\mathrm{oneloop},\mathrm{CT}}
 = 
 T_R N_Q
 \left\{
  \frac{2 \left( p^2 g^{\mu\nu} - p^\mu p^\nu \right)}{D_1^\flat D_2^\flat}
  \left[ 1 - \frac{4 k \cdot \left(p-p^\flat\right) + p^2}{4 D_1^\flat} + \frac{4 k \cdot \left(p-p^\flat\right) - p^2}{4 D_2^\flat} \right]
 \right.
 } & & 
 \nonumber \\
 & &
 \left.
 + 
  \frac{\left[ 8 k^\mu k^\nu-2g^{\mu\nu} \left( D_1^\flat + D_2^\flat \right) \right]}{D_1^\flat D_2^\flat}
  \left[ 
        1 - \frac{4 k \cdot \left(p-p^\flat\right) + p^2}{4 D_1^\flat} + \frac{4 k \cdot \left(p-p^\flat\right) - p^2}{4 D_2^\flat} 
 \right. \right.
 \nonumber \\
 & &
 \left.\left.
        + \left( k \cdot \left(p-p^\flat\right) \right)^2 \left( \frac{1}{\left(D_1^\flat\right)^2} + \frac{1}{\left(D_2^\flat\right)^2} - \frac{1}{D_1^\flat D_2^\flat}\right)
  \right]
 - \frac{p^2 g^{\mu\nu}}{D_1^\flat D_2^\flat}
 \right.
 \nonumber \\
%
%
 & &
 \left.
 - \frac{3}{14} \frac{\left( p^\flat \cdot \left(p-p^\flat\right) \right)^2 p^\flat{}^\mu p^\flat{}^\nu }{\left(D_1^\flat\right)^2 \left(D_2^\flat\right)^2}
 + \left[
          \left( \frac{1}{3} p^\flat \cdot \left(p-p^\flat\right) - \frac{p^2}{2} \right) \left(p-p^\flat\right)^\mu \left(p-p^\flat\right)^\nu
 \right. \right.
 \nonumber \\
 & & 
 \left. \left.
        + \left( \frac{2}{15} p^\flat \cdot \left(p-p^\flat\right)  - \frac{p^2}{2} \right) 
          \left( \left(p-p^\flat\right)^\mu p^\flat{}^\nu + p^\flat{}^\mu \left(p-p^\flat\right)^\nu \right) 
        - \frac{1}{6} \left(p-p^\flat\right)^2 p^\flat{}^\mu p^\flat{}^\nu
 \right. \right.
 \nonumber \\
 & & 
 \left. \left.
        + \frac{2}{5} \left(p^\flat \cdot \left(p-p^\flat\right)\right)^2 g^{\mu\nu}
        + \frac{1}{6} \left( \left(p-p^\flat\right)^2 + 2p^2 \right) p^2 g^{\mu\nu}
        - \frac{4}{15} p^2 p^\flat{}^\mu p^\flat{}^\nu
   \right] 
   \frac{D_1^\flat+D_2^\flat}{\left(D_1^\flat\right)^2 \left(D_2^\flat\right)^2}
 \right\}.
 \nonumber \\
\eq
The terms in the first three lines approximate $R_{\mathrm{oneloop}}$ in the on-shell and in the ultraviolet
limit.
The terms in the fourth to sixth line ensure that the integration of $R_{\mathrm{oneloop},\mathrm{CT}}$
gives the desired result.
We have
\bq
 g^2 \mu^{2\eps} S_\eps^{-1}
 \int \frac{d^Dk}{\left(2\pi\right)^D}
 R_{\mathrm{oneloop},\mathrm{CT}}
 & = &
 \frac{\alpha_s}{4\pi} 
 \;
 i Z_{3,Q}^{(1)} \left( - g^{\mu\nu} p^2 + p^\mu p^\nu \right).
\eq
Let us note that the last term
\bq
   - \frac{4}{15} T_R N_Q p^2 p^\flat{}^\mu p^\flat{}^\nu
   \frac{D_1^\flat+D_2^\flat}{\left(D_1^\flat\right)^2 \left(D_2^\flat\right)^2}
\eq
only vanishes linearly in the on-shell limit.
It is however proportional to $p^\flat{}^\mu p^\flat{}^\nu$ and will give a vanishing contribution when contracted
into quantities, which vanish when contracted with $p^\flat{}^\mu$ or $p^\flat{}^\nu$.


\section{Conclusions}
\label{sect:conclusions}

In this paper we showed that residues (or cuts) from raised propagators can be made to vanish for renormalised quantities in the on-shell scheme.
This is a significant simplification for numerical methods at two-loops and beyond.
We achieve this by constructing an integral representation for the ultraviolet counterterms in the on-shell scheme.
We worked out these counterterms explicitly for $\phi^3$-theory and QCD.

\subsection*{Acknowledgements}

This work has been supported by the 
Cluster of Excellence ``Precision Physics, Fundamental Interactions, and Structure of Matter'' 
(PRISMA+ EXC 2118/1) funded by the German Research Foundation (DFG) 
within the German Excellence Strategy (Project ID 39083149).


\begin{appendix}

\section{Feynman rules}
\label{sect:Feynman_rules}

In this appendix we list the Feynman rules for $\phi^3$-theory.
The Feynman rule for the propagator is
\bq
\begin{picture}(85,20)(0,5)
 \Line(70,10)(20,10)
\end{picture} 
 & = &
 \frac{i}{p^2-m^2+i\delta},
 \nonumber \\
\eq
with $\delta$ an infinitesimal small positive number.
The vertex is given by
\bq
\begin{picture}(100,35)(0,50)
\Vertex(50,50){2}
\Line(50,50)(80,50)
\Line(50,50)(29,71)
\Line(29,29)(50,50)
\end{picture}
 \;\; = \;\;
 i \lambda^{(D)}.
 \\ \nonumber
\eq
The coupling $\lambda^{(D)}$ is defined in eq.~(\ref{def_lambda_D}).
The Feynman rules for the counterterms are
\bq
\begin{picture}(85,20)(0,5)
 \Line(70,10)(20,10)
 \Line(40,5)(50,15)
 \Line(40,15)(50,5)
\end{picture} 
 & = &
 i \left[ \left(Z_\phi-1\right) p^2 - \left( Z_\phi Z_m^2 - 1 \right) m^2 \right],
 \nonumber \\
\begin{picture}(100,35)(0,50)
\Line(56,50)(80,50)
\Line(43,57)(29,71)
\Line(29,29)(43,43)
\Line(45,45)(55,55)
\Line(45,55)(55,45)
\end{picture}
& = &
 i \left(Z_\phi^{\frac{3}{2}} Z_\lambda-1\right) \lambda^{(D)}.
 \\ \nonumber
\eq

\end{appendix}

{\footnotesize
\bibliography{/home/stefanw/notes/biblio}
\bibliographystyle{/home/stefanw/latex-style/h-physrev5}
}

\end{document}